# CLDTA: Contrastive Learning based on Diagonal Transformer Autoencoder for Cross-Dataset EEG Emotion Recognition

Yuan Liao, Yuhong Zhang, Shenghuan Wang, Xiruo Zhang, Yiling Zhang, Wei Chen, Yuzhe Gu, and Liya Huang*

*Abstract*— Recent advances in non-invasive EEG technology have broadened its application in emotion recognition, yielding a multitude of related datasets. Yet, deep learning models struggle to generalize across these datasets due to variations in acquisition equipment and emotional stimulus materials. To address the pressing need for a universal model that fluidly accommodates diverse EEG dataset formats and bridges the gap between laboratory and real-world data, we introduce a novel deep learning framework: the Contrastive Learning based Diagonal Transformer Autoencoder (CLDTA), tailored for EEG-based emotion recognition. The CLDTA employs a diagonal masking strategy within its encoder to extracts full-channel EEG data's brain network knowledge, facilitating transferability to the datasets with fewer channels. And an information separation mechanism improves model interpretability by enabling straightforward visualization of brain networks. The CLDTA framework employs contrastive learning to distill subject-independent emotional representations and uses a calibration prediction process to enable rapid adaptation of the model to new subjects with minimal samples, achieving accurate emotion recognition. Our analysis across the SEED, SEED-IV, SEED-V, and DEAP datasets highlights CLDTA's consistent performance and proficiency in detecting both task-specific and general features of EEG signals related to emotions, underscoring its potential to revolutionize emotion recognition research.

## I. INTRODUCTION

Recently, the domains of human-computer interaction and affective computing have seen substantial advancements due to the exploration of emotion recognition [1]. Compared to facial expressions, movements, and linguistic cues, electroencephalogram (EEG) provides a more direct and objective measurement of human emotional responses [2], [3]. It boasts high temporal resolution and is difficult to fake or conceal [4], [5]. Moreover, in contrast to other neural imaging modalities such as fMRI and ECOG, EEG is non-invasive and relatively easy to collect, which has led to an increasing focus on EEG-based emotion recognition in the field of Brain-Computer Interface [6], [7], [8]. This interest has spurred the development and launch of affordable, gel-free wireless EEG devices, alongside studies employing a few electrodes for detecting emotions and depression [9], [10], [11]. In these studies, deep learning methods have achieved significant results in EEG-based emotion recognition. However, a pressing issue is that current deep learning models, despite performing well on various datasets, need different parameters set for each (e.g., SEED [12] and DEAP [13]) to accommodate distinct EEG data configurations. This laborious and time-consuming retraining process significantly hampers progress in emotion recognition using EEG in real-world scenarios.

Extensive research has been conducted on the EEG representation of emotions [14], [15]. Current emotion recognition methodologies widely apply Differential Entropy (DE) features [16]. It has also been found that analyzing functional connectivity is essential for the advancement of emotion recognition. Based on this, graph neural networks (GNNs) [17] and convolutional neural networks (CNNs) [18] have been proposed to extract spatial embedding of DE features among different EEG channels. Furthermore, long short-term memory (LSTM) [19] and attention mechanisms have been utilized to learn emotion-related EEG representations [20]. These approaches leverage the end-to-end capabilities of deep neural networks, thereby eliminating the need for manual feature extraction.

Though effective, these methodologies heavily rely on two key assumptions: data quality and quantity. Firstly, the quality of data from consumer-grade EEG devices often falls short compared to that from strictly controlled laboratory environments, resulting in few models performing well across both dataset types [21]. Secondly, popular models such as GNNs and CNNs are typically designed for specific EEG datasets, which prevents current deep learning models from acting as a bridge to connect multiple datasets and facilitate the sharing of knowledge [17], [22]. Moreover, due to subject variability, current research primarily focuses on collecting extensive data for each participant and conducting lengthy

This work was supported by the National Natural Science Foundation of China (Grant No. 61977039), "New Infrastructure Development & University Informatization" research project of China Association for Educational Technology (Grant No. XJJ1202205007) and Open subject of cognitive EEG and transcranial, electrical stimulation regulation of neuracle (Grant No. BRKOT-NJUPT-20220630H). (*Corresponding author: Liya Huang*).

Yuan Liao, Xiruo Zhang, Yiling Zhang, Wei Chen, Yuzhe Gu and Liya Huang are with college of electronic and optical engineering & college of flexible electronics (future technology), Nanjing University of Posts and Telecommunications, Jiangsu, 210023, China, E-mail: {1022020619; 1022020620;1022020621;1222025223;1222025429; huangly}@njupt.edu.cn.

Yuhong Zhang is with the Department of Bioengineering, University of California, San Diego, La Jolla, 92093, USA, email: yuz291@ucsd.edu.

Shenghuan Wang is with the College of Letters and Science, UC Davis, Davis, CA 95618, USA (e-mail: dvswang@ucdavis.edu).

training [3] to learn emotion patterns based on individual-specific EEG representations [9], [13], necessitating periodic model recalibration to ensure stable accuracy.

The scenario highlights the value of transfer learning (TL) for EEG-based emotion recognition, facilitating improvement through knowledge transfer. Key contributions include the BiDANN model by Li et al. [23], focusing on generalizability and feature identification; the RGNN by Zhong et al. [24], aimed at cross-subject variation and noise reduction; and the PR-PL framework by Zhou et al. [25], designed for more accurate, individualized recognition by minimizing label dependence. These TL approaches address core challenges in affective BCI, demonstrating significant advancements in emotion detection from EEG signals. The primary issue with existing approaches is their focus on applying transfer learning within a single dataset without checking if the learned features work across various datasets. Furthermore, these approaches typically use all data from the target domain[26], meaning all EEG records for new subjects must be available before transferring knowledge. This is impractical in real-world situations where data may be limited and quick emotion detection is necessary.

We introduce a novel contrastive pretraining transfer learning framework, named Contrastive Learning based on Diagonal Transformer Autoencoder (CLDTA), to enhance the performance of emotion recognition from EEG data in real-world scenarios. Our approach draws inspiration from BERT's Masked Language Modeling (MLM), where we simulate full-channel laboratory data and limited-channel real-world data by masking portions of EEG channels. Diagonal masking strategy and information separation technique trains the model to identify emotional representations that are independent of the number of EEG channels, thereby improving the model's applicability across various data settings. Furthermore, by comparing EEG signal samples from the same or different subjects, our model uncovers more generalizable emotional representations, independent of the subjects [22]. Knowledge transfer is then applied to utilize the learned emotional features and model parameters in real-world emotion recognition tasks.

The CLDTA framework is structured into two main stages: the contrastive learning process and the calibration-prediction process. Initially, the Diagonal Transformer Autoencoder (DTA) learns to represent emotions from EEG signals. Contrastive learning is then employed to amplify the alignment of features corresponding to identical emotions and to diminish the alignment of features corresponding to differing emotions. Subsequently, in the calibration-prediction stage, the pre-trained DTA, coupled with a newly initialized classifier, is fine-tuned using a small set of labeled samples from new subjects, ensuring swift personalization. Post-calibration, the model is equipped to perform precise emotion classification from EEG data. The integration of diverse data augmentation techniques during the contrastive learning phase significantly enhances the model's robustness and applicability across various datasets.

In summarize, the CLDTA model offers several distinct advantages:

- **Universality:** By implementing a diagonal masking strategy, the model can effectively learn brain network knowledge from high-quality, full-channel EEG datasets and apply it to realistic, relatively noisy, and lesser-channel EEG datasets. This enhances the model's universality, making it suitable for various data acquisition devices and aBCI use scenarios.
- **Rapid Adaptation:** By integrating contrasting learning and transfer learning mechanisms, the model rapidly adjusts to new subjects with minimal samples. This swift adaptation eliminates the need for extensive training, providing significant benefits for real-time emotion recognition applications.
- **Interpretability:** As a result of an information separation mechanism, the model converts EEG signals into understandable structures, facilitating visualization and analysis of individual emotion attributes. This innovation helps mitigate the common 'black-box' issue associated with deep learning models.
- **Validated Effectiveness:** Due to various data augmentation mechanisms, the robustness and accuracy of CLDTA have been validated across four publicly available EEG datasets: SEED, SEED-IV [27], SEED-V [28].

The structure of this paper is organized as follows: Section II outlines the related work, providing context and background for our study. Section III describes the methodology, including the development and implementation of our model. Section IV details the experimental setup, data collection, and evaluation metrics. Section V presents the results and offers an analysis of the findings. Finally, Section VI concludes the paper with a discussion of the implications, limitations, and future directions for research in this area.

## II. RELATED WORKS

### A. EEG-based Emotion Recognition

EEG-based emotion recognition involves feature extraction and classification, traditionally leveraging discrete wavelet transform (DWT), power spectral density (PSD), differential entropy (DE), and differential asymmetry (DASM) with SVM or LDA classifier [29].

Compared to conventional machine learning algorithms, deep learning has introduced end-to-end approaches that autonomously extract features using CNN, LSTM, GNN. For instance, Wang et al. [30] proposed a self-supervised EEG emotion recognition model based on CNN to enhance resource utilization efficiency. Ma et al. [31] developed a multimodal residual LSTM (MM-ResLSTM) network, while Song et al. [17] proposed a dynamic graph convolutional neural network (DGCNN) for EEG emotion recognition. More recently, the advent of Transformer models [32] achieved significant success in fields such as Natural Language Processing and Computer Vision. The emergence of Transformers also represents a significant evolution in discerning emotional states. Wang et al. [33], used attention mechanisms to focus on key features, helping to classify emotions by combining data from different parts of the brain.

*B. Transfer Learning in EEG Processing*

The high variability in individual EEG signals [34] limits the generalizability of deep learning methods in emotion recognition, confining many models to lab settings despite potential wider applications [35]. Transfer learning, aimed at applying knowledge from one domain to another, has shown promise in EEG analysis, especially in cross-session, cross-subject, and cross-database scenarios [36]. Research has primarily focused on cross-session and cross-subject scenarios to mitigate EEG signal variability over time and between individuals. Zhang et al. [37] introduced a similarity-guided transfer learning method using Maximum Mean Discrepancy (MMD) and TrAdaBoost for closer data distribution alignment. Domain adaptation (DA) techniques like the bi-hemispheres domain-adversarial neural network (Bi-DANN) [23] and regularized graph neural network (RGNN) aim to learn domain-invariant representations. Domain generalization (DG) methods, such as the two-phase prototypical contrastive domain generalization framework (PCDG) [38] and the Contrastive Learning method for Inter-Subject Alignment (CLISA) [22], reduce reliance on new subject data by identifying subject-invariant emotional representations. Li et al. [15] proposed a graph-based multi-task self-supervised learning model (GMSS) for more general representation learning.

In cross-database scenarios, addressing differences between databases remains challenging but crucial for model adaptability. Lin et al. [39] developed a personalized model using robust principal component analysis (RPCA) to reduce intra- and inter-individual differences. Wang et al. [40] analyzed electrode-frequency distribution maps (EFDMs) with CNNs, noting high-frequency bands' effectiveness in emotion recognition. Liu et al. [41] introduced CD-EmotionNet, a transfer learning-based model for enhancing emotion recognition with few-channel EEG data, marking a step towards cross-device adaptability.

III. METHODOLOGY

*A. Overall Framework*

This section introduces our Contrastive Learning based on the Diagonal Transformer Autoencoder (CLDTA). As illustrated in Fig. 1, the architecture encompasses both the pre-training procedure of contrastive learning and the calibration-prediction process in emotion recognition. The pre-training phase of CLDTA involves five key components: data preprocessing and feature extraction, augmentation, the DTA Encoder, the projector, and the contrastive loss function. Initially, samples are drawn from the EEG data bank and then processed and feature extracted followed by generating a broader sample range through the data augmentation module. The DTA Encoder subsequently extracts emotion features based on brain networks from each EEG signal. Ultimately, the projector maps the properties into a high-dimensional feature space to compute the contrastive loss, optimizing the DTA Encoder and projector. During the calibration-prediction phase, the model, which integrates the pre-trained DTA Encoder and an initialized classifier, is fine-tuned using a small set of labeled samples from new subjects. This step enables accurate emotion detection in new subjects. Once calibrated, the model is then ready for emotion recognition tasks.

*B. Data preprocess*

The initial step in our process is to preprocess EEG signals to yield high-quality, artifact-free data. To obtain a more relevant and lower-dimensional representation for emotion recognition, we utilize the widely-used differential entropy (DE) feature, which is defined as follows:

$$DE(X) = -\int f(x)\log f(x)dx = \frac{1}{2}\log(2\pi e\sigma^2) \quad (1)$$

where $\sigma^2$ is the variance of the signal. Differential entropy features of each segment were extracted separately in the $\delta$ (0.1-4 Hz), $\theta$ (4-8 Hz), $\alpha$ (8-13 Hz), $\beta$ (13-31 Hz), and $\gamma$ (31-50 Hz) frequency bands. In one experiment of a subject, the DE features trained from continuous samples across time are concatenated and smoothed with a linear dynamic system (LDS) model [12].

*C. Data Augmentation*

Data augmentation enhances our model by diversifying data representation and acting as a regularizer to improve robustness and performance. We have adopted effective augmentation techniques for DE data, specifically MixUp method [42] and Masking technique [43], after thorough evaluation.

(1) MixUp

MixUp facilitates the model's ability to discern shared information among positive pairs of samples. The MixUp data augmentation process creates a new sample, by linearly combining a pair of randomly selected training samples, $x_i$ and $x_j$ as follows:

$$x' = \lambda' x_i + (1 - \lambda')x_j \quad (2)$$

where $\lambda$ is a value sampled from a Beta distribution.

(2) Masking

The masking technique, otherwise referred to as channels dropout, has been demonstrated to yield superior results with sizable training sets [44]. This data augmentation method applies a mask that sets a random subset of channels to zero, introducing controlled noise and distortion. This procedure can be mathematically expressed as:

$$x'_i = x_i * mask \quad (3)$$

where 'mask' is a vector of zeroes and ones of length 62.

*D. DTA Encoder*

This section introduces the Diagonal Transformer Autoencoder (DTA), as depicted in Fig. 2. It draws on the fundamental principles of the Transformer encoder [32].

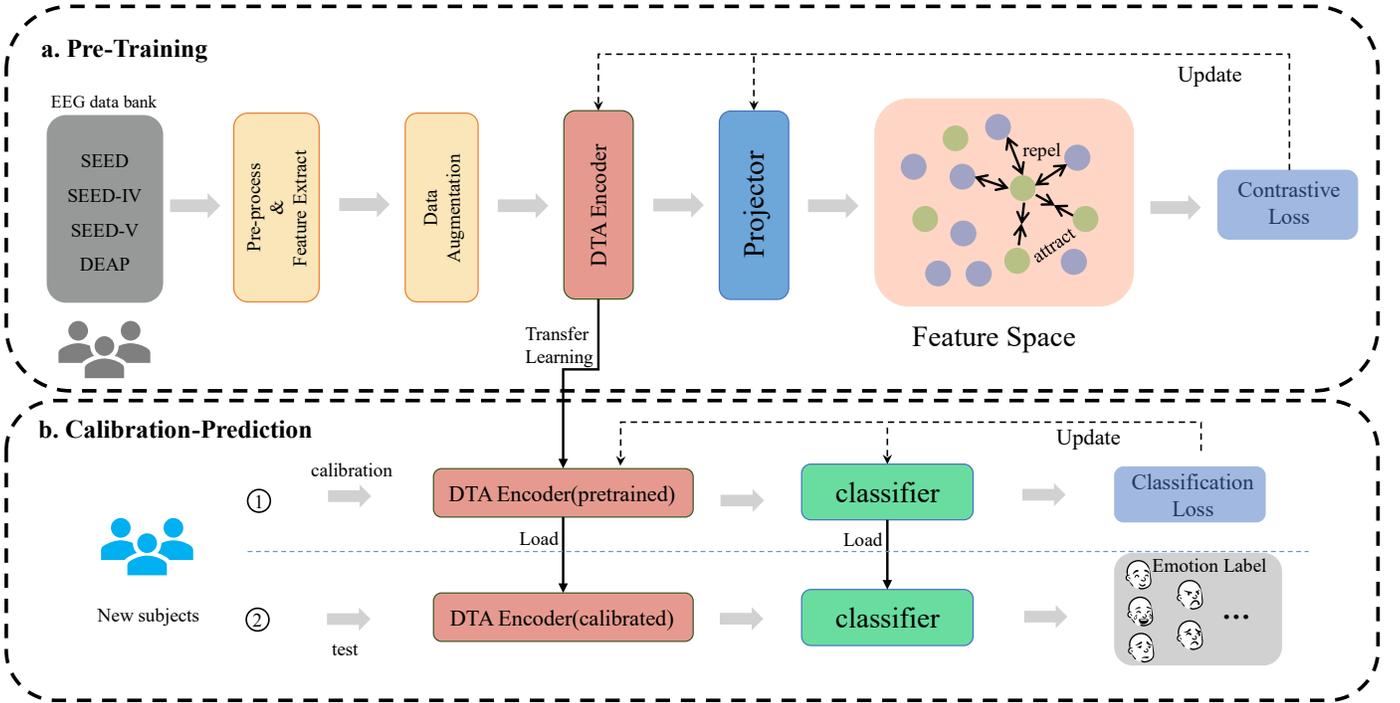

**Fig. 1.** Overview of the Transfer Learning Pipeline Using Contrastive Learning based on Diagonal Transformer Autoencoder (CLDTA). (a) Pre-Training: This phase involves preparing the EEG data from SEED, SEED-IV, SEED-V, and DEAP datasets, which undergoes preprocessing and data augmentation before being fed into the DTA Encoder. The encoder's output is then projected, and the model is updated based on contrastive loss, which aims to cluster similar emotion features closer in the feature space while pushing dissimilar ones apart, as indicated by the "attract" and "repel" arrows among subjects' representations. (b) Calibration-Prediction: This phase consists of two steps. First, a small subset of labeled samples from a new subject is collected to calibrate the pre-trained DTA Encoder and classifier. Next, the calibrated DTA Encoder and classifier are then utilized for subsequent emotion recognition in the same subject. Calibration adjusts the model to the new subject's EEG for better accuracy and the classifier links features to emotions for predicting the subject's emotional state.

Following the approach in [45], we incorporate a diagonal masking strategy (highlighted in blue in Fig. 2) to extract brain network knowledge, effectively bridging the gap between full-channel EEG data and fewer-channel EEG datasets. Moreover, we use an information separation mechanism (indicated by orange dashed lines in Fig. 2) to isolate the learned knowledge, thereby enhancing the model's interpretability.

(1) Diagonal Masking Strategy

The self-attention mechanism tends to assign excessively high attention weights to nodes themselves, as shown in Fig. 3(a). When processing EEG data with fewer number of channels, this sparsity of information can result in diminished accuracy, depicted in Fig. 3(b). To counteract this issue, Fig. 3(c) reveals that, in the pre-training stage, we capitalize on the Transformer's high parallel processing capability to focus on learning the brain network knowledge provided by full-channel EEG datasets through the Diagonal Masking strategy.

The attention mechanism of the Transformer consists of three vectors query, key and value (QKV). V is updated based on the matching degree of Q and K (i.e., attention matrix $A$). In this process, Diagonal Masking Operation is like $A_{ii} = 0$.

$$Diagmask(A) = (J - I) * A \qquad (4)$$

where $J$ represents a full matrix of ones, and $I$ represents an identity matrix.

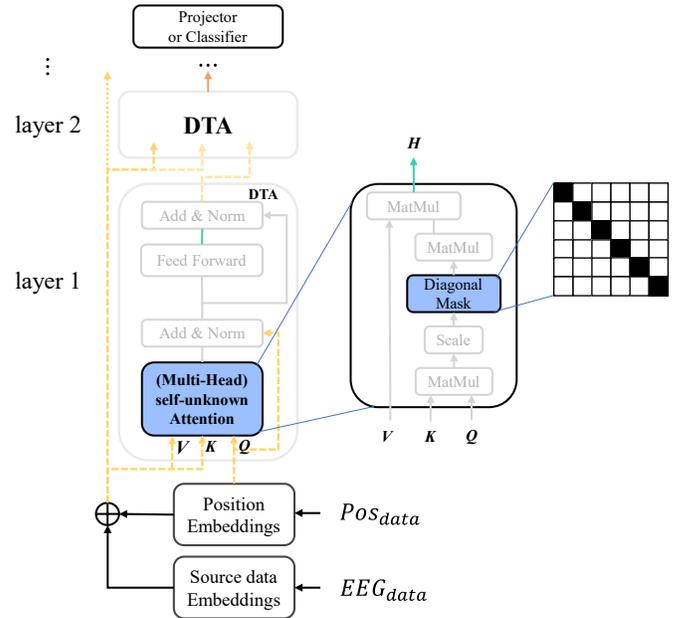

**Fig. 2:** Architecture of the DTA. The blue box signifies the diagonal masking strategy, and the orange dashed arrows represent the information separation mechanism, marking adaptations from the standard Transformer design.

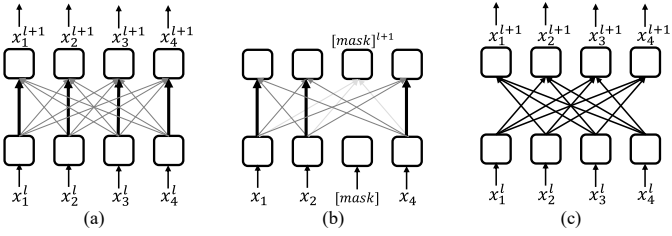

**Fig. 3.** Visualization of Self-Attention Weights and Diagonal Masking strategy in EEG Channel Analysis. Line thickness and shade indicate connection strength and influence. (a) Predominant self-attention in node weighting. (b) Self-Attention Weights after masking node 3 data. (c) Enhanced learning brain network knowledge through Diagonal Masking in pre-training.

(2) Information Separation Mechanism

The Transformer architecture uses residual connections, which impacts the Diagonal Masking Operation's self-unknown-attention abilities. Furthermore, the connections between EEG channels only can be understood by examining attention weights. An input separation method has been implemented to address this issue. Through this mechanism, the key value input (KV) for each encoding layer is isolated from the network flow and is fixed as a combination of the input encoding and position encoding. The query input (Q) is the only component that gets updated across layers. Fig. 2 illustrates this information isolation mechanism within the DTA, indicated by orange dashed arrows.

(3) Position Embedding and Source Data Embedding

The query (Q) input employs position embedding to transform the 3D coordinates of 62 nodes, derived from the 10-20 System, into the $d_{model}$ dimension using nonlinear mapping, thereby integrates prior spatial knowledge ($P_{emb}$) into the model. To augment the position encoding's expressiveness, we incorporate learnable position encoding ($L_{emb}$). Analogously, source data embedding encodes the differential entropy (DE) features to conform to the $d_{model}$ dimension size. The encoding formula is as follows:

$$P_{emb} = f_2(acvtivate(f_1(pos_{data}))) \quad (5)$$
$$S_{emb} = f_4(acvtivate(f_3(DE))) \quad (6)$$
$$Q^1 = P_{emb} + L_{emb} \quad (7)$$
$$K^1 = V^1 = Q^1 + S_{emb} \quad (8)$$

In the above formula, $pos_{data}$ represents the three-dimensional coordinates of the channel, $P_{emb}$ is the a priori position encoding. $R_{emb}$ is the learnable position embedding and $S_{emb}$ is the source data embedding. $f_i(\cdot)$ is the linear function, and acvtivate $(\cdot)$ is the activation function.

(4) Self-unknown Attention

As shown in Fig. 2, we have two encoding inputs, Source data Embedding and Position Embeddings, which are, respectively, $X = [X_1, \cdots, X_n]^T \in R^{n \times d_{model}}$ and $P = [P_1, \cdots, P_n]^T \in R^{n \times d_{model}}$, where $d_{model}$ is the feature dimension and $n$ is the number of channels. $K$ and $V$ are fixed in all layers, while the query $Q^i$ is updated with each layer.

Considering the input $Q^i K^i V^i$ of the i-th encoding layer and the output $H^i$, the formula is as follows:

$$H^i = \begin{cases} SUA(Q^i, K, V) = Softmax\left(Diagmask\left(\frac{Q^i K^T}{\sqrt{d}}\right)\right)V, if\ train \\ SA(Q^i, K, V) = Softmax\left(\frac{Q^i K^T}{\sqrt{d}}\right)V \quad\quad , if\ test \end{cases}$$
(9)

where $SUA(\cdot)$ represents the self-unknown attention layer, $SA(\cdot)$ represents the self attention layer and $H = [H_1^i, \cdots, H_m^i, \cdots, H_n^i]$.

In summary, while retaining the basic structure of the Transformer, the flow of information between the encoding layers in the CLDTA encoding module is as follows:

$$g(x) = Norm(Add(x, FeedForward(x))) \quad (10)$$
$$Q^{i+1} = DTA(Q^i, K, V) = g(norm(Add(Q^i, H^i))) \quad (11)$$

Here, we note that in the inference training process of CLDTA, the i-th element represented by $Q_i$ will not directly see the corresponding encoding representation from $Q_i^1 = P^1$ in any layer. However, during the testing phase, the diagonal masking mechanism is shut down, restoring it to self-attention.

*E. The Projector*

The nonlinear projector can help the basic encoder better learn the representation of downstream prediction tasks[46]. Here, we only use the Multilayer Perceptron (MLP), the formula is as follows:

$$Z = Projector(Q) \quad (12)$$

As shown in Fig. 4, the Projector mainly includes three fully connected layers with the number of hidden units decreasing sequentially from 128, 256, to 128. The corresponding positions in the figure show Batch Normalization, ELU and Dropout.

*F. The Contrastive Loss*

To measure the similarity of emotion-related features between two sets of samples, we can calculate the cosine similarity of the encoded representation vectors. The input batch samples $G^A = [G_1^A, \cdots, G_n^A]$ and $G^B = [G_1^B, \cdots, G_n^B]$ are transformed into $Z^A$ and $Z^B$ through the DTA encoder and the projector, respectively. Then, we can compute the cosine similarity of the feature sets between $Z^A$ and $Z^B$:

$$s(z_i^A, z_i^B) = \frac{z_i^A \cdot z_i^B}{\|z_i^A\|\|z_i^B\|}, s(z_i^A, z_i^B) \in [0,1] \quad (13)$$

The purpose of contrastive loss is to maximize the similarity of the EEG signals within the positive pair as fully as possible. We adopt the normalized temperature-scaled binary cross-entropy with logits loss computed by

$$x_i = \frac{s(z_i^A, z_i^B)}{\tau} \quad (14)$$
$$\sigma(x) = \frac{1}{1 + \exp(-x)} \quad (15)$$
$$l_n = -[y_n \cdot \log \sigma(x_n) + (1 - y_n) \cdot \log(1 - \sigma(x_n))] \quad (16)$$
$$loss = \frac{1}{N}(l_1, \cdots, l_N) \quad (17)$$

where $\tau$ is the temperature parameter for softmax. The variable y can take the values of 0 or 1 and $\sigma(x)$ denotes the sigmoid function.

The smaller the loss function, the more similar the samples in the same category and the more dissimilar the samples in different categories. Adopting this loss function allows a sample to be similar to multiple samples at the same time, thereby accelerating the training.

*G. Calibration-prediction Process*

In the calibration-prediction process, we use the pre-trained DTA encoder to extract emotional features and predict emotional labels from the representations. We optimize the parameters of the pre-trained model and classifier using the cross-entropy loss function.

The classifier is utilized to predict emotional labels from the representations extracted from the DTA encoder. As depicted in Fig. 4, the classifier primarily comprises two fully connected layers.

$$Label = Classifier\ (DTA_{pretrained}(DE, pos_{data}))\quad(18)$$

Finally, when the loss function converges, it can be used for subsequent emotional recognition of the subjects.

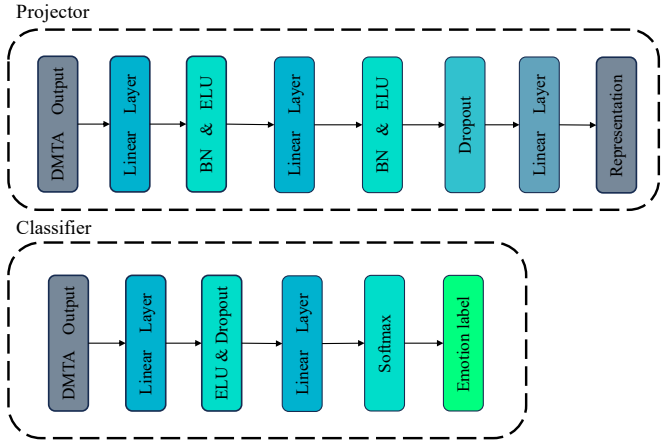

**Fig. 4.** The architecture of the Projector and Classifier. BN represent Batch Normalization. Linear layer represents fully-connected layer. ELU represents Exponential Linear Unit.

## IV. Experiments

In this section, we outline the datasets employed, elucidate the data preprocessing procedures, and expound upon the implementation details of the model. Subsequently, We define the evaluation procedures and introduce the advanced deep learning benchmarks used for comparison. Lastly, we discuss the methodologies deployed for analyzing the performance of our model.

*A. Dataset*

We first outline the datasets selected for this study and the rationale behind their selection:
(1) SEED Dataset: Developed by Zheng and Lu [12], this dataset includes EEG data from 15 subjects who watched 15 Chinese film clips, eliciting three emotions: positive, negative, and neutral. Each subject participated in three sessions, watching one clip per session for a total of 15 trials.
(2) SEED-IV Dataset: Introduced in [27], this dataset features EEG and eye movement data from 15 subjects (7 men and 8 women) responding to 72 film scenes depicting four emotions: joy, sorrow, neutrality, and anxiety. Subjects participated in three sessions with 24 trials each at different times.
(3) SEED-V Dataset: First utilized in [28], it comprises EEG and eye movement signals related to five emotions (happiness, sadness, neutral, fear, and disgust) from 15 film clips, with 16 subjects (6 males, 10 females) participating in three sessions.
(4) DEAP Dataset: Established by Koelstra et al. [13], this dataset consists of EEG signals from 32 channels and peripheral physiological signals from 8 channels, collected from 32 participants watching 40 one-minute music videos. Participants rated the videos on arousal, appeal, likes/dislikes, dominance, and familiarity.

The SEED series dataset is expected to be an excellent benchmark for pre-training the CLDTA model, as it features a significantly larger number of subjects compared to most publicly available datasets. The datasets were collected in controlled environments to induce specific emotions using video clips, with data captured via a 62-channel ESI NeuroScan system aligned with the International 10-20 system. They offer a broad range of emotional labels for a discrete emotional modeling approach, as opposed to a valence-arousal spectrum. Contrastingly, the DEAP dataset, with its different EEG equipment, data specifications, emotional stimuli, and labeling approach, presents unique challenges for cross-dataset classification tasks. This makes it an ideal candidate for assessing the model's performance across diverse datasets.

*B. Data Preprocessing*

To ensure data consistency, we re-processed the original EEG data from the datasets. This study primarily utilized the EEGLAB toolbox [47] in MATLAB for pre-processing, which includes data input, electrode positioning, filtering, baseline correction, manual identification and removal of bad segments and channels, independent component analysis(ICA), manual exclusion of irrelevant components, and re-referencing. For the SEED, SEED-IV, and SEED-V datasets, we initially applied a band-pass filter from 0.01 to 48 Hz and a 50 Hz notch filter to eliminate noise. The criteria for rejecting bad channels are as follows: channels with a flatline duration exceeding 5 seconds; channels whose variance exceeds 4 times the standard deviation of the total channel signal; and spatially adjacent channels with a correlation less than 0.6. The criteria for rejecting time segments are: if the variance in each time window exceeds 7 times the variance of the current channel, the window is discarded. EEGLAB's 'spherical' interpolation algorithm is employed to interpolate channels discarded due to volume conduction effects, assigning different interpolation weights based on the proximity of surrounding nodes. ICA is subsequently applied to remove artifacts likely caused by eye movements, muscle movements, or other environmental noise, with up to 5 ICA components being removed. The data is re-referenced using a sample mean reference. We utilize the last 30 seconds of each trial to ensure the stimulated emotions are sufficiently coherent and intense.

For the DEAP dataset, we employed the same data preprocessing method. The data was first adjusted to match the 62-channel format of the SEED-series datasets, and missing channel data was filled with zeros. We adhered to the partitioning strategy outlined in [48] and [41], which converts the dataset into binary emotion recognition tasks by segmenting the valence dimension into positive/negative and the arousal dimension into high/low arousal, with the threshold for both dimensions set at 5. Thus, the processed data can be summarized as shown in TABLE I.

TABLE I
Summary of Preprocessed Dataset Details

| Dataset | Subject | Session | Trial | Sample | Total |
|---|---|---|---|---|---|
| SEED | 15 | 3 | 15 | 30 | 20250 |
| SEED-IV | 15 | 3 | 24 | 30 | 32400 |
| SEED-V | 15 | 3 | 15 | 30 | 21150 |
| DEAP | 32 | 1 | 40 | 30 | 38400 |

Note: Trial: refers to the number of trials selected. Sample: represents the number of samples per each trial. Total: signifies the total sample count for each dataset. For the SEED-V dataset, the first trail of data from subject 5 is missing.

*C. Training Details*

We trained our CLDTA on NVIDIA RTX 3080ti GPU, pretraining the model on SEED, IV, and V datasets. The CLDTA was configured to 4 layers, model dimension ($d_{model}$) to 32, hidden layer dimension to 64, and multi-head attention count to 4. The Projector flattens the data and maps it to 128 dimensions. The temperature hyperparameter $\tau$ for contrastive learning was set to 0.5.

For optimizing the contrastive learning model, we used the Adam optimizer [49], with the initial learning rate set to 1e-4, and weight decay set to 0.005 based on empirical standards. A random seed of 42 was set, batch size was configured to 256, epoch was set to 30, dropout was set to 0.1, and activation function was set as Exponential Linear Units (ELU) [50].

For the calibrating and transfer process of emotion recognition in MLP classifier, we used two hidden layers, each with 32 units. ELU were used between every two layers. We used cross-entropy loss and Adam optimizer for parameter optimization. The learning rate was empirically set to 1e-5. Batch size was empirically set to 128. We trained for 100 epochs with early stopping (maximal tolerance of 20 epochs without validation accuracy increase).

*D. Test and Validation*

We applied the leave-one-subject-out cross-validation (LOSOCV) method to assess our approach. In LOSOCV, each subject's data is alternately used for transfer learning, with the rest for training. For each test, an equal number of labeled samples per category is selected from the target subject's test set, excluding all other unlabeled samples from training. This process repeats for all subjects' data.

Subject-dependent experiments use a small set of labeled samples from target subjects for transfer learning, with the remaining data for accuracy testing. The training and testing set division follows protocols from [41] and [12]. For SEED, training involves the first 9 trials per session, with the next 6 trials for testing. SEED-IV uses the first 16 trials for training and the last 8 for testing. SEED-V employs a triple cross-validation (10 for training, 5 for testing) for five emotion tasks. DEAP uses an 80% training and 20% testing split per subject.

In strictly subject-independent experiments, when no target subject calibration samples are available, calibration uses source subjects' data, followed by testing on the target subjects.

*E. Performance Comparison*

To investigate the effectiveness of our contrast learning method, we compared it with several notable emotion recognition methods, including A-LSTM [51], DGCNN [17], BiDANN [23], SSL-EEG [52], RGNN [24], GMSS [15], and PR-PL [25]. These methods are emblematic of prior research in emotion recognition. Their results were either directly quoted or replicated from the literature to ensure a reliable comparison with our proposed method. It's important to note that our results are compared only with advanced models under the same standard experimental settings. In our performance comparison protocols, results reproduced by our team are marked with an asterisk (*).

*F. Methods for Analyzing Model Performance*

(1) Model Stability and Channel Reduction

In the SEED series dataset, with its 62 channels from various brain locations, the excessive number of channels not only raises computational demands but also hampers the practicality of aBCI systems. Hence, it's essential to minimize channel use while analyzing EEG data. Our model calibration tests involved randomly masking EEG channels to assess the impact of channel quantity on recognition accuracy.

(2) Identifying Brain Regions for Emotion Recognition

EEG channels correspond to different brain cortex areas, each associated with specific physiological functions. To pinpoint crucial regions or channels for emotion recognition, we analyzed location encoding data. Calculating the cosine similarity between channels helped us identify the importance of nodes and their community groupings.

(3) Contrastive Learning Evaluation

We evaluated the impact of contrastive learning by visualizing features before and after encoding and by measuring inter-class divergence (ICD) and intra-class similarity (ICS). ICD evaluates the similarity level among samples of the same class in the embedding space, while ICS assesses the separation degree between different class samples. A smaller intra-class distance implies higher intra-class similarity, and a larger inter-class distance indicates greater separation.

$$CD = E[\|f(x) - f(y)\|_2^\alpha], \quad \alpha > 0 \text{ and } (x,y) \sim p_{pos} \quad (19)$$
$$ICS = E[\|f(x) - f(y)\|_2^\alpha], \quad \alpha > 0 \text{ and } (x,y) \sim p_{neg} \quad (20)$$

In these calculations, $p_{pos}$ denotes scenarios where the labels of the pair are matching, while $p_{neg}$ refers to scenarios where the labels do not match.

## V. RESULT AND DISCUSSION

Drawing from the analysis presented in Section IV, Part A, we first evaluate the CLDTA model's performance in subject-dependent and subject-independent setups on the SEED series datasets. Then, we evaluate the cross-device and cross-dataset classification tasks on the DEAP dataset.

### A. Emotion Recognition Performance on the SEED series Dataset

(1) Subject-dependent Evaluation

Three configurations of DTA were tested: DTA without transfer learning (DTA w/o TF), DTA with transfer learning within the same dataset (DTA-Single-Dataset), and DTA with transfer learning across multiple datasets (DTA-Multi-Dataset). The experimental results are shown in Table II. The results underscore the benefits of transfer learning, especially when applied across datasets, in improving the model's effectiveness. In subject-dependent evaluations, DTA shows competitive or superior performance compared to advanced models like GMSS, achieving an accuracy of 95.09% on the SEED dataset and the highest accuracy on SEED-IV and SEED-V, indicating its capability to learn stable subject features.

(2) Subject-independent Evaluation

In the subject-independent experiments detailed in Table III, it is evident that the CLDTA model outperforms the SVM baseline by achieving respective performance enhancements of 32.4%, 23.8%, and 29.9% on the SEED, SEED-IV, and SEED-V datasets. Furthermore, the CLDTA model attains state-of-the-art results on SEED-IV and SEED-V, with accuracies of 64.11% and 61.45%, respectively.

Furthermore, CLDTA consistently presents a notably low accuracy standard deviation in both testing scenarios, demonstrating its strong discrimination and generalization abilities. This comprehensive performance across different testing conditions confirms the effectiveness of the proposed transfer learning strategy in optimizing network performance, highlighting CLDTA as a viable approach for practical emotion recognition applications. However, its performance on the SEED dataset did not reach the most advanced level, which may be attributed to the dataset's broad emotional categories (positive, negative, neutral) as opposed to the more granular labels found in SEED-IV and SEED-V. These findings suggest that the efficacy of the model's learning is influenced by the granularity of emotion labeling.

TABLE II
EEG Emotion Recognition: Comparison with State-of-the-Art Methods on SEED, SEED-IV, SEED-V (Mean and SD%)

| MODEL | Dataset | | | | | |
|---|---|---|---|---|---|---|
| | SEED | | SEED-IV | | SEED-V | |
| | Acc. | Std. | Acc. | Std. | Acc. | Std. |
| SVM[53] | 83.99 | 9.27 | 56.61 | 20.05 | 69.5 | 10.28 |
| A-LSTM[51] | 88.61 | 10.16 | 69.50 | 15.65 | - | - |
| DGCNN[17] | 90.4 | 8.49 | 65.97 | 15.03 | - | - |
| BiDANN[23] | 92.38 | 7.04 | 70.29 | 12.63 | - | - |
| SSL-EEG[52] | 83.32 | 9.20 | 63.59 | 19.82 | - | - |
| RGNN[24] | 94.24 | 5.95 | 79.37 | 10.54 | - | - |
| GMSS[15] | **96.48** | 4.63 | 86.37 | 11.45 | - | - |
| PR-PL[25] | 94.84 | 9.16 | 83.33 | 10.61 | - | - |
| DTA w/o TF | 90.44 | 8.49 | 81.88 | 13.29 | 77.92 | 11.17 |
| DTA-Single-Dataset | 93.12 | 5.02 | 82.12 | 6.52 | 78.33 | 9.61 |
| **DTA-Multi-Dataset** | **95.09** | **4.48** | **88.3** | **4.62** | **80.15** | **8.33** |

— indicates the experiment results are not reported on that dataset.

Table III
Subject-Independent Accuracies on SEED, SEED-IV, SEED-V (Mean and SD%)

| MODEL | Dataset | | | | | |
|---|---|---|---|---|---|---|
| | SEED | | SEED-IV | | SEED-V | |
| | Acc. | Std. | Acc. | Std. | Acc. | Std. |
| SVM[53] | 56.73 | 16.29 | 51.78 | 12.85 | 47.3 | 16.53 |
| A-LSTM[51] | 72.18 | 10.85 | 55.03 | 09.28 | - | - |
| DGCNN[17] | **79.95** | 09.02 | 52.82 | 09.23 | - | - |
| SSL-EEG[52] | 67.52 | 12.73 | 53.62 | 08.47 | - | - |
| GMSS[15] | 76.04 | 11.91 | 62.13 | 08.33 | - | - |
| **CLDTA** | 75.09 | **05.88** | **64.11** | **04.62** | **61.45** | 10.82 |

— indicates the experiment results are not reported on that dataset.

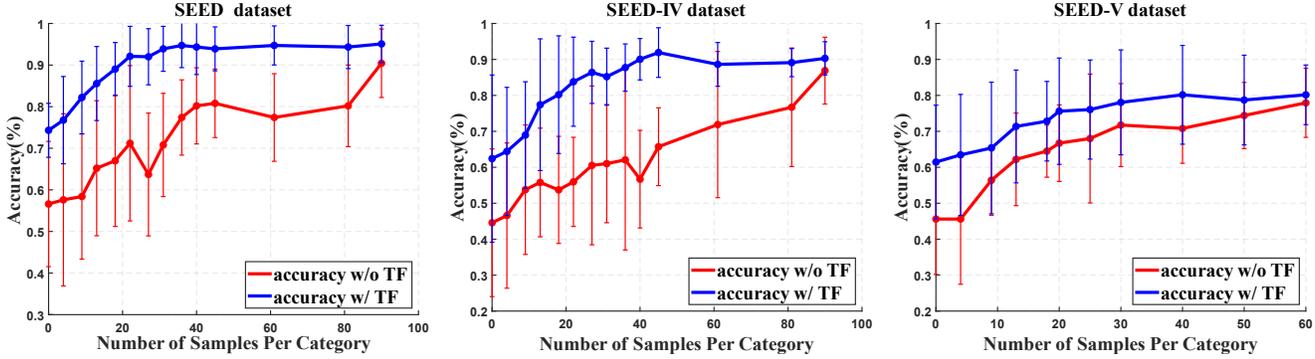

**Fig. 5.** Performance of Fully-Supervised vs. Transfer-Learning-Based CLDTA Models in Calibration Tests Across SEED Datasets: (a) SEED, (b) SEED-IV, (c) SEED-V

*B. Cross-device Cross-electrode Evaluation on the DEAP Dataset*

We compared two transfer learning strategies: one employing a model pre-trained on the SEED series datasets (SEED2DEAP) and another pre-trained on DEAP itself (DEAP2DEAP), against a baseline model with no transfer learning (Rand2DEAP). This comparison elucidates the impact of transfer learning on model efficacy in varying experimental setups. The results, as shown in Table IV, indicate that our CLDTA model, leveraging transfer learning, achieved superior accuracy. Specifically, SEED2DEAP excelled in arousal classification with a 94.11% accuracy rate and a 2.1% standard deviation, while DEAP2DEAP showed higher accuracy in Valence classification at 94.58% with a 1.4% standard deviation. These results affirm the effectiveness of our proposed transfer learning strategy in capturing cross-device and cross-electrode EEG emotion features.

TABLE IV
CLDTA vs. State-of-the-Art Methods on DEAP: Valence and Arousal Classification Accuracies

| Methods | Accuracy (Mean / SD) (%) | |
|---|---|---|
| | Valence | Arousal |
| SVM[53] | 72.59 / 9.73 | 74.44 / 9.84 |
| CD-EmotionNet [41] | 86.29 / 9.71 | 84.16 / 10.86 |
| DGCNN[17] | 86.32 / 6.04 | 83.68 / 5.68 |
| Rand2DEAP | 81.92 / 3.53 | 83.37 / 4.33 |
| SEED2DEAP | 93.31 / 1.80 | **94.11 / 2.10** |
| DEAP2DEAP | **94.58 / 1.40** | 92.58 / 1.80 |

*C. Calibration Test*

To assess performance with limited labeled samples, we explored how different quantities of labeled samples affect model calibration, comparing models with and without transfer learning. Figure 5 illustrates that when employing different quantities of calibration samples for fine-tuning, the accuracy of the CLDTA model markedly surpasses that of the fully-supervised baseline across the entire range of sample sizes, with the most pronounced advantage observed in scenarios with limited labeled data. Specifically, for the SEED dataset, CLDTA's performance nearly matches full-supervised training (90.44%) with over 20 labeled samples per category. For SEED-IV, CLDTA reaches 87.88% of full-supervised training with more than 32 labeled samples per category. For SEED-V, CLDTA's performance is close to full-supervised training (77.92%) with over 13 labeled samples per category. Beyond 40 calibration labels per category, the performance of all pre-training models converges.

In addition, we also recorded the time consumed by the model during the calibration prediction phase, as shown in Table V. The number of training iterations required for calibrating the pre-trained model is 17% of that required by the randomly initialized model. In terms of training time, this represents a time saving of 91.48%. This demonstrates that pre-trained models are both faster and more stable in calibration compared to fully-supervised models.

For the DEAP dataset, calibration tests were conducted on three models: SEED2DEAP, DEAP2DEAP, and Rand2DEAP. Results shown in Fig. 6 indicate that pre-trained models on SEED and DEAP achieve nearly similar performances with limited samples, with a mean accuracy difference of 3.7%. This indicates that CLDTA effectively captures subject-invariant emotional traits, showing resilience to differences in channel numbers and device types. The baseline model showed a higher tendency for overfitting compared to the pre-trained models, which adapted better.

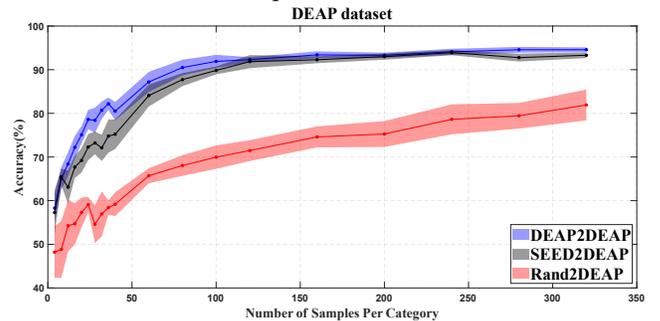

**Fig. 6.** Calibration Test Results on DEAP Dataset: Comparing SEED2DEAP, DEAP2DEAP, and Rand2DEAP Model Performances.

TABLE V
Comparison of Calibration Training Convergence between Pre-trained and Randomly Initialized Models

| Model | Epochs | Times(seconds) |
|---|---|---|
| Randomly Initialized DTA | 1084 | 636.28 |
| Pre-trained DTA | 187 | 54.23 |

*D. Stability Analysis*

For practical applications, minimizing the number of electrodes is advantageous for both feasibility and user comfort. Challenges including disconnections due to head movements, short circuits from excessive conductive paste, and potential electrode malfunctions can impair model performance. To evaluate our model's resilience in the face of such issues, we simulated real-world conditions such as electrode failure and noise interference.

We used the model pre-trained on the SEED series dataset for our experiments. In the electrode failure test, we simulated failures by setting channel data to zero or replacing it with data from nearby channels, with the number of failed channels ranging from 1 to 40. In the noise interference test, we added Gaussian noise with intensity varying from 0.1 to 3 times the sample variance. The results, shown in Fig.7, indicate the pre-trained model's superior anti-interference capability compared to a fully supervised model. A small number of electrode failures slightly improved performance by 1.21%, suggesting that redundant channels may introduce noise in emotion recognition tasks. Performance declines in the pre-trained model when failures exceed 26 electrodes, while the fully supervised model's performance gradually decreases with more failures.

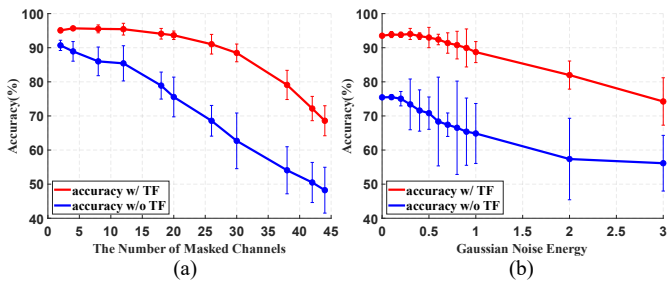

**Fig. 7.** Evaluating Pre-trained and Fully-Supervised Models' Resilience to Electrode Failure and Noise Interference. (a) Electrode failure experiment, (b) Noise interference experiments.

During the noise interference experiment, the pre-trained model consistently outperformed the fully supervised model at low noise levels. When the noise intensity was under 1, the transfer model's performance decreased by only 4%, a minor reduction compared to the 10% drop in the fully supervised model. However, as noise intensity increased from 1 to 3 times the variance, the accuracy of both models dropped—the pre-trained model by 14.5% and the fully-supervised model by 8.7%. The pre-trained model's initial stability may be due to its reliance on sophisticated features learned during pre-training, making it more resistant to minor disturbances. Yet, high noise levels impact the pre-trained model more as it may inaccurately associate enhanced noise with previously learned features, leading to performance drops. Conversely, the fully-supervised model adapts better to high noise levels, possibly because it continuously fine-tunes parameters to accommodate all variations, including noise.

*E. Explainability and Connectivity analysis*

To investigate the role of different brain regions in emotion recognition, we conducted a connection analysis after the model stabilized, focusing on the 10-20 system. By computing cosine similarity between node positions to form an adjacency matrix and retaining only connections exceeding the mean plus 1.8 standard deviations, our analysis (Fig. 8) highlights significant involvement of the frontal and temporal lobes in emotion processing, along with observed asymmetry in brain hemisphere activities. These findings align with previous studies [52],[54],[55], suggesting a correlation with the spatial distribution of emotions and activation of frontal-parietal networks in response to emotional stimuli. This underscores the distinct EEG signal characteristics during emotion recognition and suggests potential for future research using advanced graph theory to further elucidate the complex interactions between brain regions and emotions.

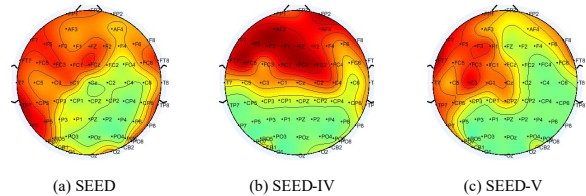

(a) SEED  (b) SEED-IV  (c) SEED-V

**Fig. 8.** Degree centrality of brain connectivity learned by the CLDTA, with darker colors indicating stronger connections to other regions.

*F. Feature Space Visualization*

This section describes the extraction of features by the CLDTA model that align between subjects, particularly when encountering new participants. We utilize the t-SNE algorithm to evaluate the model's performance on previously unseen subjects. Due to space limitations and for clarity, we randomly chose three participants and visualized their spatial characteristics for both positive and negative emotions. Figure 9 illustrates the feature distributions of these three subjects from the SEED dataset in a two-dimensional space using t-SNE.

We analyzed the ICD and ICS metrics, as detailed in equations (19) and (20), with results displayed in Table VI. Initial observations from Fig. 9(a) and (b) highlight subject variability; the same emotions across different subjects are widely spaced, often overlapping with different emotions from other subjects, underscoring the challenge of cross-subject recognition. Conversely, Fig. 9(c) shows that the CLDTA model effectively blends data across subjects while maintaining distinct emotion categories. This result indicates a substantial reduction in subject-specific features, as the model projects subject features into an emotion-centric space that is independent of the individual without needing subject-specific calibration. It suggests that pre-training via contrastive learning effectively reduces subject variability while maintaining the distinction of emotions, thereby enabling cross-subject emotion recognition. Post-calibration, as depicted in Fig. 9(d), both metrics exhibit further enhancement, demonstrating the model's capability to quickly adjust to new subjects with minimal labels, thereby markedly improving both performance and user experience.

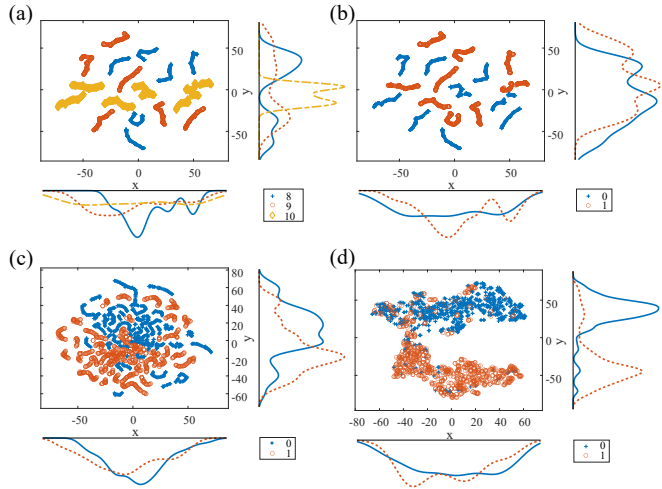

**Fig. 9.** Visualization of latent features using t-SNE on the SEED dataset. (a) t-SNE results of original differential entropy (DE) features for subjects 8, 9, 10, colored according to different subjects. (b) Original features represented through color-coding for positive and negative emotions. (c) Features extracted by the pre-trained CLDTA model without calibration for new subjects. (d) Enhanced performance of the model following subject calibration using 20 labeled samples.

TABLE VI

Computed inter-class divergence (ICD) and intra-class similarity (ICS)Metrics

|  | ICD | ICS |
|---|---|---|
| before encode: (a), (b) | 9.876 | 0.1127 |
| after encode: (c) | 37.49 | 0.0569 |
| after calibration: (d) | 73.75 | 0.0323 |

*G. Ablation study*

To examine the role of data augmentation in enhancing our model's EEG emotion recognition capability, we conducted an ablation study with the CLDTA model. This study assessed the effects of data augmentation techniques on model performance, as detailed in Table VII. Utilizing a singular augmentation approach, CLDTA with Masking (CLDTA w/ Mask) demonstrated superior results, underscoring the mask's significant contribution to improving EEG emotion signal discernibility. While CLDTA with MixUp (CLDTA w/ MixUp) didn't achieve as high accuracy as masking, it exhibited lower variance, suggesting MixUp's effectiveness in enhancing sample continuity and adaptability to new subjects. Notably, combining both augmentation methods further increased accuracy and reduced variance, indicating their complementary benefits in aiding the model's learning of distinct emotional representations.

TABLE VII

Ablation study: subject-dependent classification accuracy (mean/std) on SEED, SEED-IV, and SEED-V

| Method | Accuracy (Mean / SD) (%) | | |
|---|---|---|---|
|  | SEED | SEED-IV | SEED-V |
| CLDTA w/ MixUp | 88.21/5.27 | 83.65/8.05 | 76.50/8.28 |
| CLDTA w/ Mask | 92.50/7.63 | 84.43/9.57 | 78.80/9.65 |
| CLDTA w/ both | 95.09/4.48 | 88.30/4.62 | 80.15/8.3 |

## VI. CONCLUSIONS

This paper introduces a Transfer Learning framework utilizing contrastively pre-trained CLDTA, which encode EEG signals into subject-independent emotional representations, regardless of channel count. We tested our model against four prominent emotional databases, SEED, SEED-IV, SEED-V, and DEAP, comparing it with current benchmarks. Our CLDTA model presents several advantages over existing emotion recognition methods. It dynamically leverages spatial characteristics of EEG channels based on the 10-20 system, enabling it to accommodate diverse emotion datasets with varying channel counts. Through contrastive learning, the model potentially uncovers shared temporal-spatial patterns among different emotion categories, offering insights with neurophysiological significance. Moreover, CLDTA's ability to model new subjects with fewer calibration data and its enhanced anti-interference capabilities reduce the reliance on costly label collection and manual feature extraction. This facilitates broader and quicker deployment of emotion recognition systems, improving their practical applicability.

However, the primary training data source is the SEED series, and despite employing multiple data augmentation techniques, the limited diversity could impact the robustness and generalizability of aBCI models in real-world applications. Additionally, further discussion and research on the topic of negative transfer remain necessary. Lastly, the practical deployment of the model, particularly in environments with a lower signal-to-noise ratio than laboratory conditions, has yet to be tested.

To enhance applicability in real-world scenarios, future work will aim to collect a more diverse dataset covering a wider range of ages and scenarios, explore the feasibility of large-scale emotional BCI models, and plan to achieve high performance with fewer EEG channels. It is only when aBCI can provide stable and effective performance across sessions, subjects, and dataset tasks that it can be expected to manage the complex and varied emotional recognition scenarios in real-life.

ACKNOWLEDGMENT

The authors extend their appreciation to Xiangkai Qiu and Shenglin Wang for their invaluable input on manuscript composition and figure design. Gratitude is also due to Professor Bao-Liang Lu's team and the BCMI Laboratory for their swift provision of the SEED-related dataset and code.